\documentstyle[12pt,epsf]{article}
\newcommand{\be}{\begin{eqnarray}}
\newcommand{\ee}{\end{eqnarray}}

{
{

\def\0n{0\nu\beta\beta}

\begin{document}
\thispagestyle{empty}
\mbox{}
\vspace{0.25in}
\begin{center}
{\Large \bf Neutrinoless double beta decay with scalar bilinears \\}
\vspace{0.7in}
{\bf H.V. Klapdor-Kleingrothaus$^1$ and Utpal Sarkar$^2$\\}
\vspace{0.2in}
{$^1$ \sl Max-Planck-Institut f\"ur Kernphysik,
P.O. 10 39 80, }\\ 
{\sl D-69029 Heidelberg, Germany} \\ 
\vspace{0.1in}
{$^2$ \sl Physical Research Laboratory, Ahmedabad 380 009, India \\}
\vspace{0.7cm}
\end{center}

\vspace{2.5cm}

\begin{abstract}

One possible probe to physics beyond the standard model is to 
look for scalar bilinears, which couple to two fermions of
the standard model. We point out that the scalar bilinears
allow new diagrams contributing to the neutrinoless double
beta decay. The upper bound on the neutrinoless double beta 
decay lifetime would then give new constraints on the ratio
of the masses of these scalars to their couplings to the
fermions. 

\end{abstract}

\vspace{1cm}
\newpage

In the standard model both baryon $B$ and lepton $L$ numbers 
are conserved. The baryon number conservation gives stability
of the proton, while the lepton number conservation does not
allow any Majorana mass of the neutrino. But in recent times
non-vanishing neutrino mass has been observed in the
atmospheric neutrinos \cite{atm,sol}. The smallness of the observed mass
is naturally explained if the neutrinos are Majorana 
particles. More recently a direct evidence of lepton 
number has been announced in the neutrinoless double 
beta decay ($0 \nu \beta \beta$) \cite{ndbex}. 
This observation has several interesting
consequences \cite{ndb1,ndb2}.

The neutrinoless double beta decay constrains the neutrino
mass matrices, when combined with the atmospheric, solar
and laboratory neutrino results. The amount of neutrino 
dark matter is also limited by this observation. In addition
to these direct consequences, there are also several 
indirect consequences of the neutrinoless double beta
decay, which depend only on the upper bound of its lifetime.

\begin{figure}[thb]
\vskip 0in
\epsfxsize=100mm
\centerline{\epsfbox{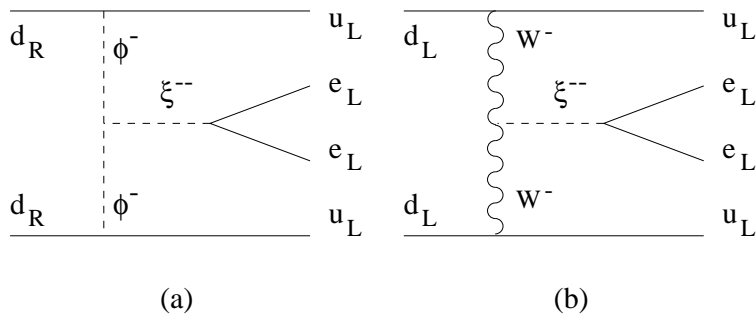}}
\vskip 0in
\caption{
Diagram for neutrinoless double beta decay with dileptons.
\label{lp0}}
\end{figure}

An upper bound on the $0 \nu \beta \beta$ decay restricts the 
scale of lepton number violation or the Majorana mass of a heavy 
right-handed neutrino \cite{nr}. In the conventional diagram, the
lepton number violation is introduced through the Majorana
mass of the neutrino. It is also possible to introduce a
doubly charged dilepton, whose couplings break lepton
number explicitly, which can then give a bound on the mass of
the doubly charged dilepton (see figure \ref{lp0}) \cite{di}. 

The doubly charged component of the $SU(2)_L$ triplet Higgs 
scalar $\xi^{--}$ allows the diagram of figure 1 with its couplings
to the standard model $SU(2)_L$ doublet Higgs $\phi$ and the leptons
$l_{iL} \equiv \pmatrix{\nu_i \cr e_i}_L$, ($i=e,\mu,\tau$), 
given by
\begin{equation}
{\cal L} = \mu \phi \phi \xi + f_{ij} l_{iL} l_{jL} \xi^\dagger .
\end{equation}
This Lagrangian can also give
a neutrino mass. In this scenario the effective neutrino mass that enters
into the $0 \nu \beta \beta$ decay is given by
\begin{equation}
<m_\nu> = \mu f_{ee} {v^2 \over m_\xi^2}
\end{equation}
where $<\phi> = v$ is the vacuum expectation value ($vev$) of 
the usual Higgs doublet $\phi$ and $m_\xi$ is the mass of the triplet
Higgs scalar $\xi$. Then the amplitude for the $0 \nu \beta \beta$
decay is given by 
\begin{equation}
A_{0 \nu}[\beta \beta] \sim \mu {m_d^2 \over v^2} f_{ee}^2 {1 
\over m_\phi^4 m_\xi^2} \sim {m_d^2 m_\nu \over v^4 m_\phi^4}. ,
\end{equation}
This contribution is much smaller than the usual contribution 
to the $0 \nu \beta \beta$ decay because of the suppression 
due to the down quark mass. A comparison of this contribution
with the present upper bound on the neutrinoless double beta 
decay lifetime can only give a very weak lower 
bound on $m_\phi$ of the order of a few GeV. 

The diagram with the $W$ boson can also give a bound on the mass of
the dileptons to its couplings. Although the down quark mass 
does not enter in the expression, now the coupling of the $W$
boson with the dilepton has a suppression proportional to the
mass of the neutrino. 
If the Higgs doublet $\phi$ is replaced by the exotic scalar
$\Sigma$, which transforms as an octet under the color
$SU(3)_c$ and doublet under $SU(2)_L$,
then this can again give a weak bound on the ratio of
the mass to the couplings of this scalar $\Sigma$. 

\begin{figure}[thb]
\vskip 0in
\epsfxsize=110mm
\centerline{\epsfbox{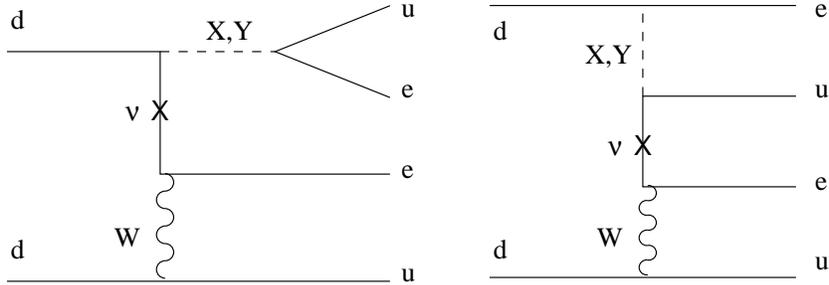}}
\vskip 0in
\caption{
Diagram for neutrinoless double beta decay involving leptoquarks.
\label{ndb-lq}}
\end{figure}

It has been argued that if there is coupling of the leptoquarks with the
usual standard model higgs doublets, then the mixing between a 
doublet leptoquark and a singlet leptoquark is allowed
\cite{lq1}. This will
then give an effective operator $d \bar u \bar \nu \bar e$, which
gives a new contribution to the $0 \nu \beta \beta$ decay,
as shown in figure \ref{ndb-lq}. From this it is then possible to 
obtain a bound on the ratio of mass and the coupling of the
leptoquark scalar \cite{lq1}, using the upper bound on the lifetime of
the $0 \nu \beta \beta$ decay. Similarly R-parity violating 
couplings are also constrained by the neutrinoless double beta
decay \cite{rp}.

Leptoquarks are only one of the possible scalars, which can
couple to a couple of fermions, In general, there are several
other scalar bilinears, which can couple to two fermions. 
In the standard model one such scalar bilinear exists,
which is the usual Higgs doublet, which couples to $\bar q_{iL} u_{R}$, 
$\bar q_{iL} d_{R}$ and $\bar \ell_{iL} e_R$ (generation
indices are suppressed). Other scalar bilinears have been
considered to understand the neutrino masses, the triplet
Higgs scalar \cite{trip} or the doubly charged dileptons \cite{zee},
which transforms under $SU(3)_c \times SU(2)_L \times U(1)_Y$
as $(1,3,-1)$ or $(1,1,-2)$ respectively. 
All possible scalar bilinears which could exist in theories
beyond the standard model have been listed
in table \ref{sc-bil}. 

\begin{table}[htb]
\begin{center}
\begin{tabular}{|c|c|c|c|c|c|}
\hline
Representation & Notation&
$qq$ & $\bar q \bar l$ & $q \bar l$ & $ll$  \\
\hline
$(1,1,-1)$ &$\chi^-$& & & & $\times$ \\
$(1,1,-2)$ &$L^{--}$& & & & $\times$  \\
$(1,3,-1)$ &$\xi$& & & & $\times$ \\
\hline
$(3^*,1,1/3)$ &$Y_a$& $\times$ & $\times$ & &  \\
$(3^*,3,1/3)$ &$Y_b$& $\times$ & $\times$ & &  \\
$(3^*,1,4/3)$ &$Y_c$& $\times$ & $\times$ & &  \\
$(3^*,1,-2/3)$ &$Y_d$& $\times$ & & &  \\
\hline
$(3,2,1/6)$ &$X_a$& & & $\times$ &  \\
$(3,2,7/6)$ &$X_b$& & & $\times$ &  \\
\hline
$(6,1,-2/3)$ &$\Delta_a$& $\times$ & & &  \\
$(6,1,1/3)$ &$\Delta_b$& $\times$ & & &  \\
$(6,1,4/3)$ &$\Delta_c$& $\times$ & & &  \\
\hline
$(6,3,1/3)$ &$\Delta_L$& $\times$ & & &   \\
$(8,2,1/2)$ &$\Sigma$& & & &  \\
\hline
\end{tabular}
\caption{Exotic scalar particles beyond the standard model.}
\end{center}
\label{sc-bil}
\end{table}

Phenomenological consequences of these scalars have been 
studied in the literature \cite{bl3,bl4}. 
The LEP constraints and the 
collider signals of these scalars have been extensively
studied \cite{bl3}. Among the various constraints, the baryon number
violating constraints and the constraints from the 
baryon asymmetry of the universe are most severe in several
cases \cite{bl4}. However, none of the present constraints conclusively
rule out the possibility of these scalars and hence further
studies to understand their phenomenology is underway. Some
of these particles are predicted to be very light in some
specific theories, which make this study even more attractive
\cite{mrs}. 

\begin{figure}[htb]
\vskip 0in
\epsfxsize=110mm
\centerline{\epsfbox{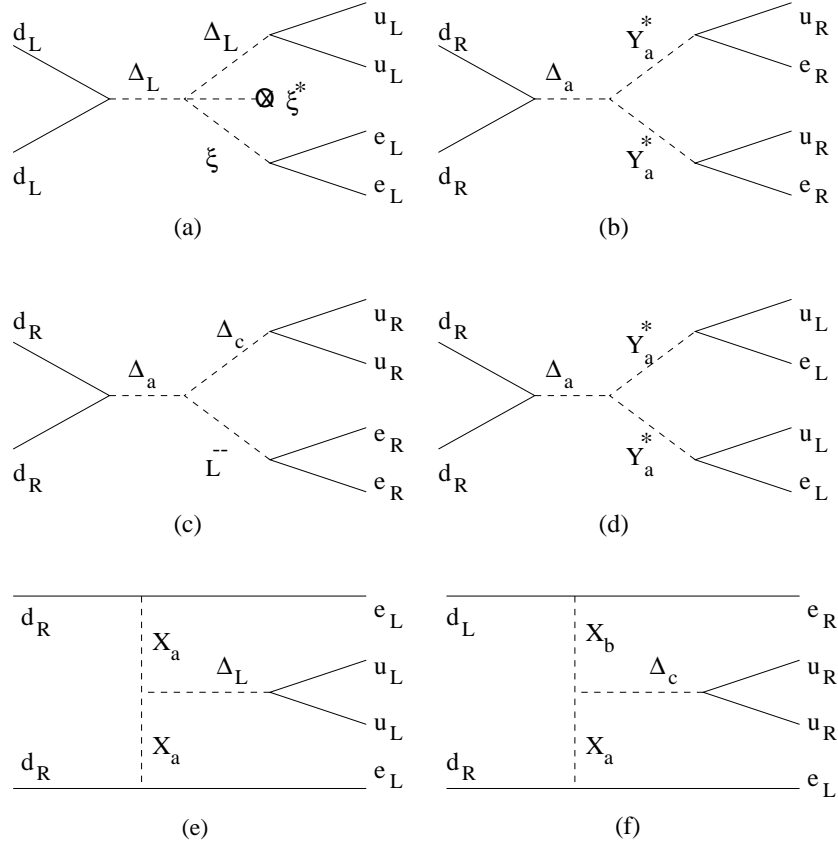}}
\vskip 0in
\caption{
New diagrams for neutrinoless double beta decay with scalar bilinears.
\label{lp2}}
\end{figure}

In this note we point out that all these scalars and their
usual couplings allow new classes of diagrams, contributing to the
$0 \nu \beta \beta$ decay. There are no standard model 
particles involved in these diagrams and the source of lepton 
number violation is the trilinear couplings of some of these
scalar bilinears. We present these diagrams in 
figure \ref{lp2}, which will now be explained. The upper
bound on the lifetime of the $0 \nu \beta \beta$ decay 
will then be used to give a bound on the ratio of the masses
and the couplings of these scalars with the usual fermions. 

We start with a systematic analysis to arrive at these diagrams.
In a $0 \nu \beta \beta$ decay process, the effective operator
is $d d \bar u \bar u \bar e \bar e$. So, we need one scalar which
couples to a $d$ quark and another $d$ or $\bar u$ or $\bar e$. The 
scalars which can couple to $d \bar u$ are $\phi$ or $\Sigma$. These
are the diagrams we already mentioned, in which a $\phi \phi \xi$
coupling gives $0 \nu \beta \beta$ decay when $\xi$ goes
to two electrons. 

Next we consider two $d$ quarks coupling to a diquark. Since both
these $d$ quarks has to be from first generation only, only those
combinations are allowed which are symmetric in generation index,
these are $\Delta_a$ and $\Delta_L$. The $Y-$type diquarks are
antisymmetric in the color indices and symmetric under $SU(2)_L$,
so they are always antisymmetric under the generation index. 

The outgoing scalars should go into a leptoquark $u e$ or 
a diquark $u u$ and a dilepton $e e$. 
Since all particles should be from the same generations, these
scalars should also be symmetric in the generation index.  
So, only the leptoquarks $Y_a$ and $Y_c$ can enter in the
process, or diquarks $\Delta_a$ or $\Delta_L$ in combination
with the dileptons $\xi$ or $L^{--}$. This consideration
allows only four possible diagrams given by figures (\ref{lp2}a
- \ref{lp2}d). 

The last possibility is when the incoming scalar is a
$d \bar e$, which are the $X-$type leptoquarks. Since they
should go into a scalar, which couples to two $u$ quarks 
symmetrically, the third scalar could be either $\Delta_c$ or
$\Delta_L$. This gives two possibilities, given by 
figures (\ref{lp2}e - \ref{lp2}f). 

In all these diagrams, there is one d i m e n s i o n f u l ???????? 
trilinear scalar coupling constant, say $\mu$, and six
powers of heavy masses in the denominator. In figure 
\ref{lp2}a, $\mu$ is of the order of $<\xi> \sim m_\nu$,
and hence the contribution of this diagram is negligible.
So, this diagram may not provide us with any new constraint.
In the remaining diagrams, $\mu$ is not related to the
electroweak symmetry breaking scale, and hence it could be
large. 

Figure \ref{lp2}b and Figure \ref{lp2}d contribute to the
$0 \nu \beta \beta$ decay by equal amount,
\begin{equation}
A_{0 \nu} [\beta \beta] = {\mu_{\Delta YY } 
f_{\Delta d d} f_{Y ue}^2
\over M_\Delta^2 M_Y^4} .
\end{equation}
The present lifetime of the 
$0 \nu \beta \beta$ decay 
\cite{ndbex}, $\tau_{1/2} = 1.5 \times 10^{25}$ yrs, 
if not fully atributed to the neutrino mass mechanism, then
gives a constraint
\begin{equation}
{\mu_{\Delta YY}f_{\Delta d d} f_{Y ue}^2 \over M_\Delta^2 M_Y^4} 
< 10^{-25} ~{\rm GeV}^{-5},
\end{equation}
considering the Fermi momentum of a nucleon inside a nucleus
to be about 200-300 MeV \cite{nucl}. 

If we assume that all the coupling constants involved in
this expression are of the order of 1, and all masses are
of the same order of magnitude, $\mu_{\Delta YY} \sim M_\Delta \sim
M_Y \sim M_s$, then we get a bound on the mass of the 
exotic scalar diquarks or the leptoquarks to be
\begin{equation}
M_s > 10^5 ~{\rm GeV} .
\end{equation}
The scalar $Y_a$ can, in general, mediate proton decay through
its couplings to two quarks or a quark and a lepton, which
gives the strongest bound on its mass to the coupling ratio, 
which is
\begin{equation}
{M_Y \over |f_{Yud} f_{Yul}|^{1/2}} > 10^{16} ~{\rm GeV}.
\end{equation}
For $f_{Yud}$ to be very small or vanishing, this bound 
is trivially satisfied and there does
not exist any bound on the $Y_a$-boson mass.
In such theories, the bound from the 
$0 \nu \beta \beta$ decay becomes the strongest bound on
the mass of the $Y_a$ and the coupling $f_{Yul}$. 

For the scalar $\Delta_a$, the strongest bound comes from
radiative proton decay, which depends on its couplings
$|f_{\Delta dd} f_{\Delta ud} f_{\Delta d \nu}|^{1/2}$. So,
if $f_{\Delta ud}$ or $f_{\Delta d \nu}$ are small, or if 
the trilinear coupling $\Delta_a \Delta_b^2$ vanishes, then
this bound goes away. In that case the strongest bound 
on $\Delta_a$ mass
comes from the $K^\circ - \overline{K^\circ}$
or $B_d^\circ - \overline{B_d^\circ}$ oscillations. The 
bounds on the
mass of $\Delta_a$ over the couplings $|f_{\Delta dd}
f_{\Delta ss}|^{1/2}$ or $|f_{\Delta dd}
f_{\Delta bb}|^{1/2}$, as obtained from the upper bound
on the lifetime of the neutrinoless double beta decay, comes
out to be $1.5 \times 10^6$ GeV and
$4.6 \times 10^5$ GeV, respectively. In case $f_{\Delta ss}$
and $f_{\Delta bb}$ are small, the strongest bound would
come from the present analysis of the $0 \nu \beta \beta$ decay.

The constraint from figure \ref{lp2}c comes out to be
\begin{equation}
{\mu_{\Delta_a \Delta_c L} f_{\Delta_a d d} f_{\Delta_c uu} f_{Lee} 
\over M_{\Delta_a}^2 M_{\Delta_c}^2 M_L^2} 
< 10^{-25} ~{\rm GeV}^{-5}.
\end{equation}
The present bound on $\Delta_c$ is much weaker, coming from
$n-\bar n$ oscillation, which depends on both $f_{\Delta_c dd}$
and $f_{\Delta_c uu}$. A slightly stronger bound $M_{\Delta_c}
/ |f_{\Delta_c uu} f_{\Delta_c cc}|^{1/2} > 10^5$ GeV
comes from $D^\circ - \overline{D^\circ}$ oscillation.
But both these bounds disappear, if the coupling of 
the $\Delta_c$ with $d$ and $c$ quarks are negligible.
Again in this case the strongest bound would come from the
$0 \nu \beta \beta$ decay. The present bounds on $L^{--}$
comes from lepton flavor changing processes, and hence involves
couplings of second and third generations of charged leptons. 
Only for the first generation, the present bound from the
$0 \nu \beta \beta$ decay becomes the strongest bound.

Figure \ref{lp2}e and \ref{lp2}f give bounds
\begin{equation}
{\mu_{\Delta_L X_a X_a} f_{X_a d e}^2 f_{\Delta_L uu}  
\over M_{\Delta_L}^2 M_{X_a}^4} < 10^{-25} ~{\rm GeV}^{-5}
\end{equation}
and 
\begin{equation}
{\mu_{\Delta_c X_a X_b} f_{X_a d e} f_{X_b de} f_{\Delta_c uu}  
\over M_{\Delta_L}^2 M_{X_a}^2 M_{X_b}^2} < 10^{-25} ~{\rm GeV}^{-5}, 
\end{equation}
respectively. The bounds
on $\Delta_L$ are similar to that of $\Delta_c$ and depends
on its couplings to $d$ and $c$ quarks.
The present bounds on $X_a$ and $X_b$ are most
stringent from the lepton flavor changing processes such as
$K^\circ \to e^+ \mu^-$, which is greater than $2.4 \times 10^5$
GeV. But these bounds involve couplings $f_{X_a s \mu}$ and
$f_{X_b s \mu}$. But the present bound is only for the first
generation, which is otherwise not constrained. 

The new contributions to the neutrinoless double beta decay
in the presence of the scalar bilinears considered here may 
have other interesting consequences. The observed 
neutrinoless double beta decay \cite{ndbex} is usually 
translated into a Majorana mass of the neutrino. 
In \cite{ma}, a specific 
model has been considered, where the observed neutrinoless
double beta decay has been explained with an almost massless
neutrinos. They considered our figure \ref{lp2}c to explain
the observed neutrinoless double beta decay with suitable
choice of parameters. There is no other source of lepton 
number violation in the model, so the neutrinos remain 
massless at the tree level. A tiny Majorana mass is then 
generated radiatively. Similar models may be constructed 
with other diagrams of figure \ref{lp2}.

To summarize, we have shown that there are new contributions
to the neutrinoless double beta decay if there are scalar 
bilinears in theories beyond the standard model. The present 
lifetime of the neutrinoless double beta
decay gives bounds on the ratio of the masses
to some of the couplings of the new scalars entering in these
diagrams. We discuss under which condition the
new bounds on these diquarks, leptoquarks
and dileptons are stronger than all available constraints.

\vskip .2in
\centerline{\bf Acknowledgement}

\vspace{0.5cm}
\noindent
One of us (US) thanks Max-Planck-Institut f\"ur Kernphysik 
for hospitality.


\newpage
\begin{thebibliography}{99}

\bibitem{atm} S. Fukuda {\it et al.}, Super-Kamiokande Collaboration, 
Phys. Rev. Lett. {\bf 85} (2000) 3999.

\bibitem{sol} S.~Fukuda {\it et al.}, Super-Kamiokande Collaboration, 
Phys. Rev. Lett. {\bf 86} (2001) 5656; Q.~R.~Ahmad 
{\it et al.}, SNO Collaboration, Phys. Rev. Lett. {\bf 87} (2001) 071301.

\bibitem{ndbex} H. V. Klapdor-Kleingrothaus {\it et al.}, 
Mod. Phys. Lett. {\bf A16} (2001) 2409; Particle and Nuclei, Letters
{\bf 110} (2002) 57; Foundations of Physics, {\bf 32} (2002) 1181.  

\bibitem{ndb1} H.V. Klapdor-Kleingrothaus, H. P\"as and A. Yu. Smirnov, 
Phys. Rev. 
{\bf D 63} (2001) 073005; {\it Dark Matter in Astro- and Particle Physics},
ed. Klapdor-Kleingrothaus H V, (Heidelberg, Springer-Verlag) (2001) 423;
H.V. Klapdor-Kleingrothaus and U. Sarkar 
Mod. Phys. Lett. {\bf A 16} (2001) 2469.

\bibitem{ndb2} 
F. Vissani, hep-ph/9708483; R. Adhikari and
G. Rajasekaran, Phys. Rev. {\bf D 61} (1999) 031301(R);
H. Georgi and S.L. Glashow, Phys. Rev. {\bf D 61} (2000) 097301;
 H.V. Klapdor-Kleingrothaus and U. Sarkar, Phys. Lett. {\bf B 532} 
(2002) 71; V. Barger, S.L. Glashow, D. Marfatia and
  K. Whisnant, Phys. Lett. {\bf B 532} (2002) 15; 
V. Barger, S.L. Glashow, P. Langacker and D. Marfatia, Phys. Lett.
{\bf B 540} (2002) 247; Y. Uehara, Phys. Lett. {\bf B 537} (2002) 256;
F. Feruglio, A. Strumia and F. Vissani, Nucl. Phys. {\bf B637} (2002) 345;
E. Ma and G. Rajasekaran, Phys. Rev. {\bf D64} (2001) 113012; 
E. Ma, Mod. Phys. Lett. {\bf A17} (2002) 289; 
Mod. Phys. Lett. {\bf A17} (2002) 627;
K. S. Babu, E. Ma, and J. W. F. Valle, hep-ph/0206292;
Z. Xing, Phys. Rev. {\bf D 65} (2002) 077302;

\bibitem{nr} M. Hirsch, H.V. Klapdor-Kleingrothaus and O. Panella, 
Phys. Lett. {\bf B 374} (1996) 7; 
H.V. Klapdor-Kleingrothaus and H. P\"{a}s, 
{\it Proc "COSMO 99": 3rd Int Conf on Particle Physics and the
Early Universe, Trieste, Italy, 1999}; hep-ph/0002109.

\bibitem{di} R.N. Mohapatra and J.D. Vergados, Phys. Rev. Lett. 
{\bf 47} (1981) 1713.

\bibitem{lq1} M. Hirsch, H.V. Klapdor-Kleingrothaus and S.G. Kovalenko
Phys. Rev. {\bf D 54} (1996) R4207; Phys. Lett. {\bf B 378} (1996) 17.

\bibitem{rp} 
M. Hirsch, H.V. Klapdor-Kleingrothaus and S.G. Kovalenko 
Phys. Rev. Lett. {\bf 75} (1995) 17; Phys. Lett. {\bf B 352} (1995) 1; 
Phys. Lett. {\bf B 403} (1997) 291; {\it Nucl. Phys. Proc. Suppl.}
{\bf A 52} (1997) 257; 
R.N. Mohapatra, Phys. Rev. {\bf D 34} (1996) 3457; 
K.S. Babu and R.N. Mohapatra, Phys. Rev. Lett. {\bf 75} (1995) 2276.

\bibitem{trip} 
G. B. Gelmini and M. Roncadelli, Phys. Lett. {\bf 99B} (1981) 411; 
C. Wetterich, Nucl. Phys. {\bf B187} (1981) 343;
 G. Lazarides, Q. Shafi, and C. Wetterich, Nucl. Phys. {\bf B181} (1981) 287;
R. N. Mohapatra and G. Senjanovic, Phys. Rev. {\bf D23} (1981) 165;
E. Ma and U. Sarkar, Phys. Rev. Lett. {\bf 80} (1998) 5716;
G. Lazarides and Q. Shafi Phys. Rev. {\bf D 58} (1998) 071702;
W. Grimus, R. Pfeiffer and T. Schwetz, Eur. Phys. J. {\bf C 13} (2000) 125;
E. Ma, T. Hambye and U. Sarkar, Nucl. Phys. {\bf B 602} (2001) 23. 

\bibitem{zee} A. Zee, Phys. Lett. {\bf B 93} (1980) 389;
L. Wolfenstein, Nucl. Phys. {\bf B 175} (1980) 93.

\bibitem{bl3} M. Leurer, Phys. Rev. {\bf D 49} (1994) 333;
S. Davidson, D. Bailey and B.A. Campbell, Zeit. Phys. {\bf C 61} (1994) 613;
F. Cuypers and S. Davidson, Eur. Phys. J. {\bf C2} (1998) 503;
J.P. Bowes, R. Foot and R.R. Volkas, Phys. Rev. {\bf D54} (1996) 6936.

\bibitem{bl4} E. Ma, M. Raidal and U. Sarkar, Eur. Phys. J. 
{\bf C 8} (1999) 301; Phys. Rev. {\bf D 60} (1999) 076005.

\bibitem{mrs} E. Ma, M. Raidal and U. Sarkar, Phys. Rev. Lett.
{\bf 85} (2000) 3769; Nucl. Phys. {\bf B 615} (2002) 313.

\bibitem{nucl} H. P\"as, M. Hirsch, S.G. Kovalenko and 
H.V. Klapdor-Kleingrothaus, Phys. Lett. {\bf B 453} (1999) 194; 
M. Hirsch, H.V. Klapdor-Kleingrothaus, S.G. Kovalenko,
Phys. Lett. {\bf B 372} (1996) 181; (E) {\bf B 381} (1996) 488. 

\bibitem{ma} E. Ma and B. Brahmachari, 
Phys. Lett. {\bf B 536} (2002) 259.

\end{thebibliography}
\end{document}